\documentclass{article}
\usepackage{graphicx}
\usepackage{epsfig}
\usepackage{amsfonts}
\usepackage{amssymb}
\usepackage{amsmath}
\usepackage[shortlabels]{enumitem}

\date{\today}

\newcommand{\insertplot}[5]{\begin{figure}
 \hfill\hbox to 0.05in{\vbox to #5in{\vfill
 \inputplot{#1}{#4}{#5}}\hfill}
 \hfill\vspace{-.1in}
 \caption{#2}\label{#3}
 \end{figure}}
 \newcommand{\inputplot}[3]{
 \special{ps: plotfile #1}
}
\newcounter{fig}

\newcommand{\be}{\begin{equation}}
\newcommand{\ee}{\end{equation}}
\newcommand{\bea}{\begin{eqnarray}}
\newcommand{\eea}{\end{eqnarray}}

\newcommand{\diff}{\mathrm{d}}

\begin{document}

 \title{ 
Hairy black holes, boson stars and non-minimal coupling to curvature invariants
} 

\author{
{\large Y. Brihaye}$^{\dagger}$, 
{\large L. Ducobu} $^{\dagger}$
\\ 
\\
$^{\dagger}${\small Physique-Math\'ematique, Universit\'e de
Mons, Mons, Belgium}
}
\maketitle 
\begin{abstract} 
The Einstein-Klein-Gordon Lagrangian is supplemented by a non-minimal coupling of the  scalar field
to specific geometric invariants~: the  Gauss-Bonnet term and the Chern-Simons term. The non-minimal coupling
is chosen as a general quadratic polynomial in the scalar field and allows -- depending on the parameters --
for large families of hairy black holes to exist. These solutions are
characterized, namely, by the number of nodes of the scalar function. The fundamental family encompasses 
black holes whose scalar hairs appear spontaneously and  solutions presenting shift-symmetric hairs.
When supplemented by a an appropriate potential, the model possesses
both hairy black holes and non-topological solitons~: boson stars.
These latter exist in the standard Einstein-Klein-Gordon equations; it is shown that
the coupling to the Gauss-Bonnet term modifies considerably  their domain of classical  stability.
\end{abstract} 
\section{Introduction}
Attempts to escape the rigidity of the minimal Einstein-Hilbert 
formulation of gravity and  the limited number of parameters 
describing its fundamental solutions -- the black holes --, lead naturally
physicists to emphasize enlarged models of  gravity. 
Besides their purely Academic interests, these attempts are largely
motivated nowadays by intriguing problems such as inflation, dark matter and dark energy.

One of the most popular class of extensions of Einstein gravity consists in
the inclusion of  scalar fields and appeals for natural interactions between the scalar fields and the 
geometry  through higher curvature terms, leaving a lot of freedom.
The general construction of scalar-tensor gravity leading to second order field equations was first obtained in  \cite{horndeski}.
Recently this theory was revived in the context of Galileon theory \cite{galileon}
and different  extensions of it, see e.g.  \cite{Deffayet:2011gz}.

Apart from their cosmological implications, the
extended models of gravity (by scalar or other types of fields) offer possibilities to escape 
the limitations of  the no-hair theorems \cite{no_hole_old,no_hole_new} holding in standard gravity.
In the last few years,  black holes endowed  by scalar hairs have attracted a lot of attention
and have  been studied in numerous theories. One particularly interesting result is the family of hairy black holes 
 constructed in \cite{Herdeiro:2014goa} within the Einstein gravity minimally coupled to a complex scalar field.
In this case,  the  no-hair theorems \cite{no_hole_old,no_hole_new} are bypassed by the
rotation of the black hole and  the synchronization of the spin of the black hole with the angular frequency of the scalar field. 
Recent reviews on the topic of black holes with scalar hairs 
can be found e.g. in \cite{Herdeiro:2015waa},\cite{Sotiriou:2015pka},\cite{Volkov:2016ehx}.

The general theory of scalar-tensor gravity \cite{horndeski}, \cite{Deffayet:2011gz} contains a lot of 
arbitrariness  and the study of compact objects such as black holes, neutron stars or boson stars needs to be 
realized  in some particular cases.
As an example, the truncation of the Galileon theory to a lagrangian 
admitting a shift-symmetric scalar field was worked out by Sotiriou and Zhou (SZ in the following)   
\cite{Sotiriou:2013qea} and still leads to a large family of models. 
Hairy black holes were constructed  perturbatively and numerically 
in the particular case of a scalar field  coupled linearly to the Gauss-Bonnet invariant \cite{Sotiriou:2014pfa}.

Abandonning the hypothesis of shift-symmetry,  several groups \cite{Doneva:2017bvd}, 
\cite{Silva:2017uqg}, \cite{Antoniou:2017acq}
considered during the past years, new types of coupling terms  between a scalar field and 
specific geometric invariants (essentially the Gauss-Bonnet term). 
In these models the occurrence of hairy black holes
results from an unstable mode of the scalar field equation in the background of a vacuum metric (the probe limit).
The interacting term of the scalar field with the curvature invariant plays a role of potential
and the coupling constant the  role of a spectral parameter. 
By continuity, the hairy black holes then exist as solutions of the full system.
It is used to say that the hairy black holes appear through a 
{\it spontaneous scalarization}   for a sufficiently large value of the coupling constant. 

In the present paper we will consider a model of scalar-tensor gravity encompassing the theories
presenting a spontaneous scalarization and the shift-symmetry property. 
Families of classical solutions whose pattern extrapolates smoothly
between shift-symmetric  hairy black holes  and  spontaneous scalarized ones will be constructed . 
The type of structure found  holds when coupling the scalar field to the 
Gauss-Bonnet invariant and to the Einstein-Chern-Simons invariant as well.
All black holes  solutions found are supported by the non-minimal coupling between 
the scalar field and the curvature invariant; however the field equations admit other types of solutions:
boson stars. These regular solutions exist with a minimal coupling of scalar field to gravity
but it will be shown that the non minimal coupling has important consequences on their stability properties.

The paper is organized as follow : in Sect. 2 we present the model to be studied. Namely the 
Einstein-Klein-Gordon Lagrangian extended by a non-minimal coupling.
We discuss the spherically symmetric ansatz and the general form of the field equations.
Sect. 3 is devoted to the presentation of the hairy black holes occurring  in the model.  
The boson stars are presented in Sect. 4 with an emphasis on the influence
of the non-minimal coupling of the spectrum of the solutions. Conclusions are drawn in Sect. 5.   
Similar results hold for Einstein-Chern-Simons gravity and are the object of the Appendix;
the activation of the Chern-Simons term is realized by means of a NUT charge \cite{NUT}.

\section{The model}

\subsection{The action}
\label{action}
We are interested in  solutions of the field equations associated with the action 
\be
   S = \int \diff^4 x \sqrt{-g} \bigg[ \frac{1}{16 \pi \mathcal{G}} R - \nabla_{\mu} \phi^* \nabla^{\mu} \phi - V(\phi) + f(\phi) {\cal I}(g)  \bigg],
\label{lagrangian}
\ee
which extends the minimal Einstein-Klein-Gordon lagrangian. Here $R$ is the Ricci scalar and $\phi$
represents a complex scalar field which -- in some circumstances -- will be chosen real.
The usual  Klein-Gordon kinetic term is supplemented by a potential $V(\phi)$ which will actually be chosen as a function of the combination $|\phi|^2 \equiv \phi\phi^*$ in order to ensure a $U(1)$ global symmetry for the scalar sector. 
In the following  $V$ will be set in the form
\be
\label{potential}
      V(\phi) = m^2 |\phi|^2 + \lambda_4 |\phi|^4 + \lambda_6 |\phi|^6
\ee
which is used generically for obtaining Q-balls in the absence of gravity and  
boson stars when gravity is set in (see e.g. \cite{jetzer}, \cite{Schunck:2003kk} for reviews).

The gravity sector is  supplemented by a non-minimal coupling between  the scalar field and the geometrical invariant ${\cal I}(g)$.
For this paper, we will be interested in the case where this invariant is the Gauss-Bonnet-scalar :
\[{\cal I}(g)={\mathcal L}_{GB}\equiv R^2 - 4 R_{ab}R^{ab}+ R_{abcd}R^{abcd}.\]
It is well known that this invariant is a total derivative in four dimensions but it will contribute non trivially 
to the equations of motion through the non-minimal coupling to the scalar field via $f(\phi)$.
For the seek of generality, we have also investigated the case of a coupling to the Chern-Simons invariant, see \ref{appendix}.

In order to preserve the $U(1)$ symmetry of the ``usual'' scalar sector, we will assume that, just like the potential, $f(\phi)$ is a function of  $|\phi|$. In this paper, we will emphasize the effects of a coupling function of the form
\be
\label{f_function}
               f(\phi) = \gamma_1 |\phi| + \gamma_2 |\phi|^2
\ee
where $\gamma_1$, $\gamma_2$ are independant coupling constants.
Several forms of the function $f(\phi)$ have been emphasized in the literature where the scalar field is 
usually choosen real. The EGB theory with  $\gamma_2 = 0$ and $V=0$ corresponds to a shift-symmetric theory
studied by SZ
 \cite{Sotiriou:2013qea}, the case $\gamma_1=0$  is considered in \cite{Doneva:2017bvd}, 
\cite{Silva:2017uqg}. 
\cite{Antoniou:2017acq}. Several choices of the function $f(\phi)$
have been  considered in  \cite{Antoniou:2017hxj} and very recently in \cite{Minamitsuji:2018xde},\cite{Silva:2018qhn}.
Solutions with the form of $f(\phi)$ above with two independant constants $\gamma_1, \gamma_2$ was, to our knowledge,
not yet investigated. 

\subsection{Equations of motion}

The equations of motion (EOM) for the general action \eqref{lagrangian} read

\begin{equation}
	\label{eqG}
	G_{\mu\nu} = 8\pi\mathcal{G} \left(T_{\mu\nu}^{(\phi)} + T_{\mu\nu}^{(\mathcal{I})}\right)
\end{equation}
for the metric function, and 
\begin{equation}
	\label{eqphi}
	- \square \phi = - \frac{\partial V}{\partial \phi^*} + \frac{\partial f}{\partial \phi^*} \mathcal{I}(g)
\end{equation}
for the scalar field.
In these equations, $G_{\mu\nu}$ is the Einstein tensor and $\square = \nabla_{\mu}\nabla^{\mu}$. The energy momentum
$T_{\mu\nu}^{(\phi)}$ arise from the variation of the standard Klein-Gordon lagrangian~:
\begin{equation}
T_{\mu\nu}^{(\phi)} = \nabla_{\left(\mu\right.}\phi \nabla_{\left.\nu\right)}\phi^* - \left(\nabla_{\alpha} \phi^* \nabla^{\alpha} \phi + V(\phi) \right) g_{\mu\nu} \  \ .
\end{equation}
Finally,
 $T_{\mu\nu}^{(\mathcal{I})}$ is the energy momentum tensor 
associated to the non-minimal coupling term\footnote{The expression of $T_{\mu\nu}^{(\mathcal{I})}$ is generically quite involved and depends on the explicit form of $\mathcal{I}(g)$. The expression of $T_{\mu\nu}^{(\mathcal{I})}$ for the case considered here can be found in \cite{Brihaye:2018bgc} with the same notations.} $f(\phi)\mathcal{I}(g)$.

From Eq.\eqref{eqphi}, one can see that the invariant $\mathcal{I}(g)$ will act as a source term for the scalar field. Consequently, if one find a space-time solution of the EOM such that $\mathcal{I}(g) \neq 0$, this solution will automatically present a non-trivial scalar field. This mechanism is known as ``curvature induced scalarization''.

\subsection{The ansatz}

\subsubsection{Metric}

We will be interested in spherically symmetric solutions. In this case, it is well known that (in the appropriate coordinate system) the metric can always be set in the form
\be
\label{metric_ss}
     \diff s^2 = - N(r)\sigma^2(r) \diff t^2 + \frac{1}{N(r)} \diff r^2 + g(r) (\diff \theta^2+\sin^2 \theta \diff\varphi^2) \ \ ,
\ee
where $\theta$ and $\varphi$ are the standard angles parameterising an $S^2$ with the usual range and $r$ and $t$ are the radial and time coordinates respectivelly.

The usual coordinate choice $g(r) = r^2$ will be used throughout this paper.

\subsubsection{Scalar field}

Within the same coordinate system, we choose a scalar field of the form

\begin{equation}
\label{scalar}
\phi(x^\mu) = e^{-i \omega t} \phi(r),
\end{equation}
where $\omega$, the frequency of the scalar field, is a real parameter and $\phi(r)$ a real function.
The scalar field will be assumed to be real (\textit{i.e.} $\omega=0$) in the case of hairy black holes.

This choice above is motivated by the construction of boson stars. 
Indeed, it is well known \cite{jetzer} that boson stars exist as solutions of the minimal Einstein-Klein-Gordon equations provided the scalar field is chosen complex (typically of the form \eqref{scalar}) and supplemented by  a mass term 
(or a more general potential \eqref{potential}) in the equations.

\subsubsection{Reduced equations}
With the Ansatz \eqref{metric_ss}--\eqref{scalar}, the equations \eqref{eqG}--\eqref{eqphi} reduces to a system of three coupled differential equations
(plus a constraint) for
the radial functions $N, \sigma$ and $\phi$. Using suitable combinations of the equations, the system is amenable to the form
\be
\label{radial_equations}
           N' = F_1( N, \sigma, \phi, \phi')\ \ , \ \ 
					\sigma' = F_2( N, \sigma, \phi, \phi') \ \ , \ \ 
					\phi'' = F_3( N, \sigma, \phi, \phi')
\ee
where $F_a$, with $ a = 1,2,3$, are involved algebraic expressions whose explicit form is not illuminating enough to 
be given.
\subsubsection{Rescaling and units}\label{rescale}
In the coming discussion we will set $c=1$ and $8\pi G=1$. The equations are then invariant under the rescaling
\be
      r \rightarrow \lambda r \ \ \ , m^2 \rightarrow \frac{m^2}{\lambda^2} \ \ \ , \ \ \   
      \lambda_{4,6} \rightarrow \frac{\lambda_{4,6}}{\lambda^2}
       \ \ \ \gamma_{1,2} \rightarrow \lambda^2  \gamma_{1,2} \ ,
\ee
where $\lambda$ has the dimension of {\it length}$^{-1}$. 
These rescaled quantities will be used in the following.
In the case of black holes we will use it to set the event horizon to unit (\textit{i.e.} $r_h=1$).
In the case of boson stars (which have no horizon) we will set the mass parameter $m$ to one ($m=1$).

\vspace{5mm}



\section{Hairy black holes}
\subsection{Boundary conditions}

We now discuss the black holes solutions of the equations. As stated above 
these solutions exist for a real scalar field, so we set $\omega = 0$ in the equations.
Let us first consider the solutions occuring   in the absence of potential   
 (\textit{i.e.} setting $m= \lambda_4  = \lambda_6 = 0$ in \eqref{potential}); the influence
of a mass term will be emphasized separately, see Sect. \ref{massterm}.

For black holes, the metric is  required to present a regular horizon at $r=r_h$, \textit{i.e.} $N(r_h)=0$.
The occurence of this condition in the equations and the requirement 
of a regular function $\phi(r)$ at the horizon  implies
a non trivial relation for the scalar function and its derivative at $r=r_h$.  
The two conditions at the horizon are summarized as follows
\be
\label{regular_condition}
       N(r_h)=0 \ \ \ , \ \ \ \phi'(r_h) = 
			\frac{-r_h^2 \pm \sqrt{\Delta} }
			  {8 r_h(\gamma_1 + 2 \gamma_2 \phi(r_h))} \ \ , \ \ 
				\Delta = r_h^4 - 96 \gamma_1^2 - 384 (\gamma_2^2 \phi(r_h)^2 + \gamma_1 \gamma_2 \phi(r_h))
\ee
 Remark that $\Delta \geq 0$ constitutes a necessary condition for solutions
to exist. We will see in the next section that it largely determines the domain of the coupling constants $\gamma_1$,$\gamma_2$
for which solutions exist.
The requirement for the solutions to be asymtotically Minkowski further implies
\be
\label{asymptotic_condition}
       \sigma(r \to \infty) = 1 \ \ \ \ , \ \ \ \ \phi(r \to \infty) = 0 \ \ .
\ee 
The four conditions  \eqref{regular_condition}-\eqref{asymptotic_condition} constitute 
the boundary values of the field equations. 
The black holes can be characterized by their mass $M$ and 
 the scalar charge $Q_s$. These are related respectively to the asymptotic decay
of the functions $N(r)$ and $\phi(r)$~:
\be
\label{mass_charge}
     N(r) = 1 - \frac{2M}{8 \pi r}  + O(1/r^2) \ \ \ , \ \ \ \phi(r) = \frac{Q_s}{r} + O(1/r^2) \ \ .
\ee
The entropy $S = \pi r_h^2$ and temperature $T_H = \sigma(r_h) N'(r_h)/(4 \pi)$ characterize the solutions at the horizon.
Using the equations the temperature can further be specified ~:
\be
\label{nprime}
     N'(r_h) = \frac{1}{r_h + 4 \phi'(r_h)(\gamma_1 + 2 \gamma_2 \phi(r_h))} \ \ .
\ee
Because the equations do not admit closed form solutions, we solved the 
system by using the numerical routine COLSYS \cite{COLSYS}
which is well adapted for the problem at hand. 
It is based on a collocation method for boundary-value differential equations and on damped Newton-Raphson iterations.
The equations are solved with a mesh of a few hundred points and  relative errors of the order of $10^{-6}$.
The values $M,Q_S, S, T_H$ can be  extracted with such an accuracy from the numerical datas.

\subsection{Numerical results}
\label{numresults}
\subsubsection{Fundamental branch}
We now present the pattern of  solutions in the $\gamma_1, \gamma_2$ parameter space.
Practically, we start from the hairy black holes constructed  in \cite{Sotiriou:2013qea},
\textit{i.e.} the shift-symmetric theory, corresponding to $\gamma_2=0$. 
A pair of solutions exist for $\gamma_1 \leq \sqrt{1/96} \approx 0.1021$ (with our convention of the non minimal coupling); 
characterized by  the sign $\pm$  appearing in the condition (\ref{regular_condition}). 
We will essentially focus on the family of solutions corresponding to the ``+'' sign which, 
in the limit $\gamma_1 \to 0$, smoothly approach the Schwarschild solution. 
Solutions corresponding to the ``--'' sign can be
constructed as well (see e.g. \cite{Brandelet:2017nbc}), forming a second branch whith higher mass.
This branch, however is difficult to  construct numerically. Moreover no regular solution can be associated to the $\gamma_1 \to 0$ limit for this branch since the value $\phi'(r_h)$ in \eqref{regular_condition} clearly diverge in this case ($\gamma_2 = 0$) for the ``$-$'' sign.
The understanding of this branch is then not aimed in the present paper.


For a fixed  value of the parameter $\gamma_1$,  the SZ solution
can be deformed by increasing (or decreasing) gradually the coupling constant $\gamma_2$. 
The pattern of hairy black holes obtained in this way turns out to be quite different for
the small values of $\gamma_1$ (say for $\gamma_1 \leq 0.005$) and for $0.005 < \gamma_1 < \sqrt{1/96}$.
For definiteness let us first discuss the family of black holes corresponding to  $ \gamma_1 > 0.005$. 

\begin{figure}[h!]
\begin{center}
{\label{non_rot_cc_1}\includegraphics[width=12cm]{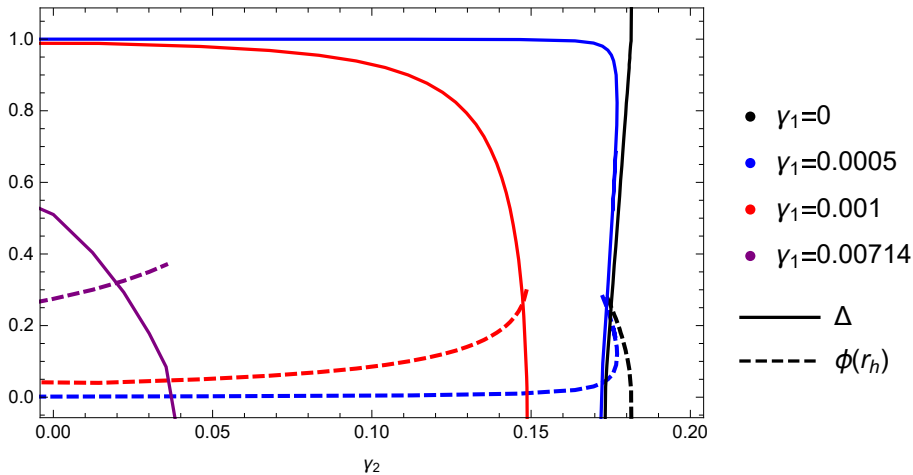}}
\end{center}
\caption{The parameter $\Delta$ (solid lines) and the value $\phi(r_h)$ (dashed lines)
 as functions of $\gamma_2$ for several values of $\gamma_1$.
\label{fig_mix_1}
}
\end{figure}

\begin{figure}[h!]
\begin{center}
{\label{non_rot_cc_1}\includegraphics[width=8cm]{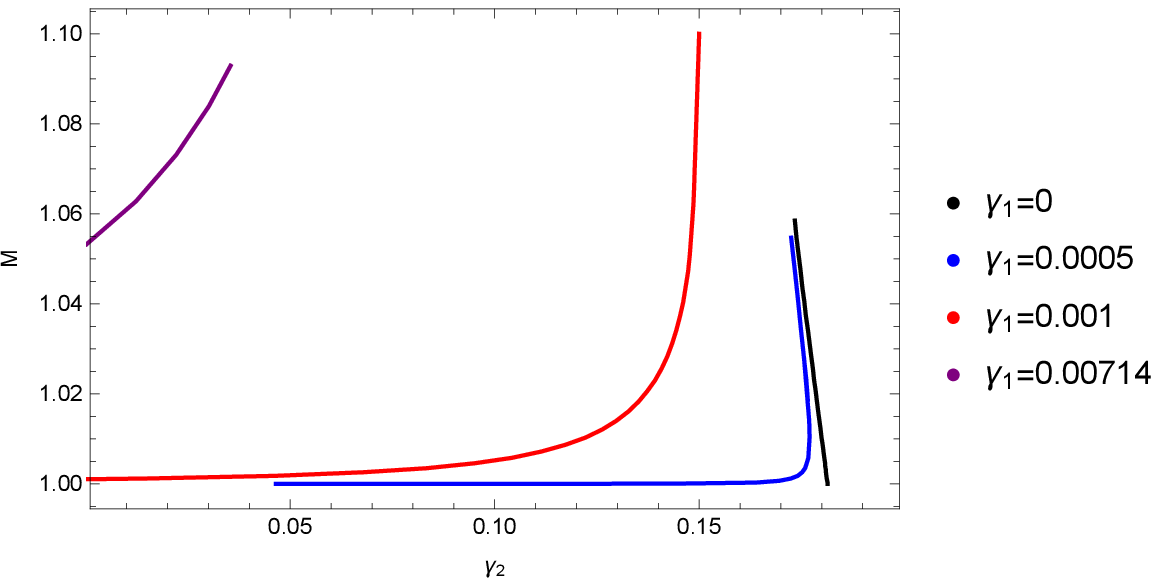}}
{\label{non_rot_cc_2}\includegraphics[width=8cm]{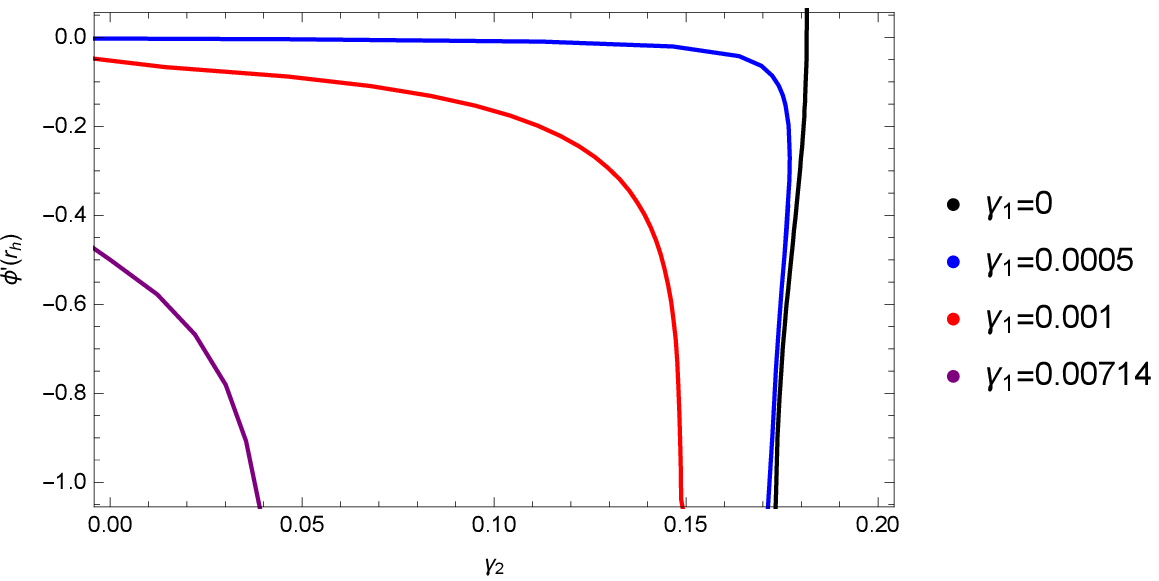}}
\end{center}
\caption{The mass  as functions of $\gamma_2$ of the solutions of Fig.1.
Right~: Idem for the value $\phi'(r_h)$.
\label{fig_mix_2}
}
\end{figure}

\begin{enumerate}[(i)]
 \item Increasing gradually the coupling constant $\gamma_2$, it turns out that
the  value $\Delta$ approaches zero at some  critical value, say  $\gamma_{2,c}$.
Accordingly, no solution exist for $\gamma_2 > \gamma_{2,c}$.
This is illustrated on Fig. \ref{fig_mix_1} where the quantities
$\Delta$  (solid lines) and $\phi(r_h)$ (dashed lines) 
are plotted as functions of $\gamma_2$ for two values of $\gamma_1$ (see the purple and red lines).
The  corresponding values of the mass and of $\phi'(r_h)$ is presented on both sides of Fig. \ref{fig_mix_2}.
\item In the case $\gamma_2 < 0$, a Schwarzschild metric can be approached arbitrarily close, although not exactly.
This is due to the fact that the scalar field never reaches $\phi(r)=0$ due to the presence  of the non-homogeneous term
in the scalar field equation. Indeed for the Schwarzschild black hole of mass $M$ we have ${\cal L}_{GB} = 48 M^2/r^6$.
\end{enumerate}

The deformation of the SZ solutions in the region $\gamma_1 \leq 0.005$
for  $\gamma_2 \neq 0$ leads to  a  richer  pattern. For a fixed value of $\gamma_1 \le 0.005$ :

\begin{enumerate}[(a)]
\item Starting from the shift-symmetric solution ($\gamma_2 = 0$) and increasing $\gamma_2 > 0$, we find that  the SZ black holes forms
a  ``first branch'' of  solutions which exists up to a maximal value, say for $\gamma_2 \leq \gamma_{2,\max{}}$.
\item Then, decreasing $\gamma_2$ from $\gamma_{2,\max{}}$, a ``second branch'' of solutions exists  for $\gamma_2 \in [\gamma_{2,c}, \gamma_{2,\max{}}]$. 
As before, the value $\gamma_{2,c}$ coincide with $\Delta = 0$ and the two branches coincide in the 
limit $\gamma_2 \to \gamma_{2,\max{}} $.
Fig. \ref{fig_mix_1} illustrates this phenomenon for $\gamma_1 = 0.0005$ (see the blue line;
in this case we find $\gamma_{2,c} \approx 0.172$ and $\gamma_{2,\max{}} \approx 0.177$).
We note that, on the interval of $\gamma_2$ where the two solutions coexist, the solution  of 
the first branch has a lower mass  than the corresponding solution on the second branch.
\item For $\gamma_2 < 0$, while decreasing $\gamma_2$,
 the black holes approach a Schwarzschild metric in the same way as point (ii) above.
\end{enumerate}

To summarize, fixing low enough values of $\gamma_1$ and varying $\gamma_2 >0$, the SZ solution deforms into a family
of hairy black holes  forming two branches which exist on  specific intervals of $\gamma_2$.
We can now emphasize how this ensemble behave when taking the limit $\gamma_1 \to 0$.
It turns out that the solutions of the first branch approach uniformly 
 the Schwarzschild black hole (irrespectively of $\gamma_2$). By contrast, the solutions of the second branch 
have a non trivial limit and approach
the set of so called ``spontaneously scalarized black holes'' for $\gamma_2 \in [\gamma_{2,c}, \gamma_{2,\max{}}]$.
These solutions were constructed directly in \cite{Doneva:2017bvd},\cite{Silva:2017uqg},\cite{Antoniou:2017acq}.
 The critical values
  $\gamma_{2,c} \approx 0.1734$ and $\gamma_{2,\max{}}  \approx 0.1814$ found in these papers fit very well with our numerical datas.
The occurence of these critical values have different explanations~:
\begin{enumerate}[(I)]
\item In the limit $\gamma_2 \to \gamma_{2,c}$, the parameter $\Delta$ (see (\ref{regular_condition})) approaches zero.
\item In the limit $\gamma_2 \to \gamma_{2,\max{}}$, the scalar hairs tends uniformly to zero. 
\end{enumerate}
The value $\gamma_{2,\max{}}$ in fact corresponds to an eigenvalue of the scalar field equation
\begin{equation}
\label{klein_gordon}
        \frac{1}{r^2} \frac{\diff}{\diff r}\left[r^2 N(r) \frac{\diff}{\diff r}\phi\right] =  \gamma_2  \frac{48 M}{r^6} \phi \ \ \ , \ \ \ N(r) = 1 - \frac{2M}{r}
\end{equation}
considered in the background of  Schwarzschild solution. 
This value reflects a tachyonic  instability of the Schwarzschild solution in the theory (\ref{action}),
opening the way for the vacuum solution to evolve into a hairy black hole.
Details about  the spectrum of this equation can be found, namely in \cite{Silva:2017uqg}, \cite{Brihaye:2018bgc}.

The question about the stability of our solutions raises naturally. In the case of the hairy black holes  
occuring by spontaneous scalarisation with a quadratic coupling to the Gauss-Bonnet term, it was shown
in Refs. \cite{Minamitsuji:2018xde}, \cite{Silva:2018qhn} that the solutions present radial instabilities
which can be removed when supplementing a quartic term in the coupling function.
The study of the stability is not aimed in this paper;
however we would like to argue about the (in)stability in the case where two solutions
coexist with different masses on the interval $[\gamma_{2,c}, \gamma_{2,\max{}}]$ (see Fig. \ref{fig_mix_2}).
It is likely that the branch with the lower mass which approaches the  Schwarzschild metric
in the limit $\gamma_1 \to 0$, is linearly
stable while the branch with the higher mass, which approaches the spontaneously scalarized solutions, 
is unstable;
this last statement being  reinforced by a continuity argument 
applied to the results of  \cite{Minamitsuji:2018xde}, \cite{Silva:2018qhn}.

\subsubsection{Excited solutions}
\begin{figure}[h!]
\begin{center}
{\label{non_rot_cc_3}\includegraphics[width=12cm]{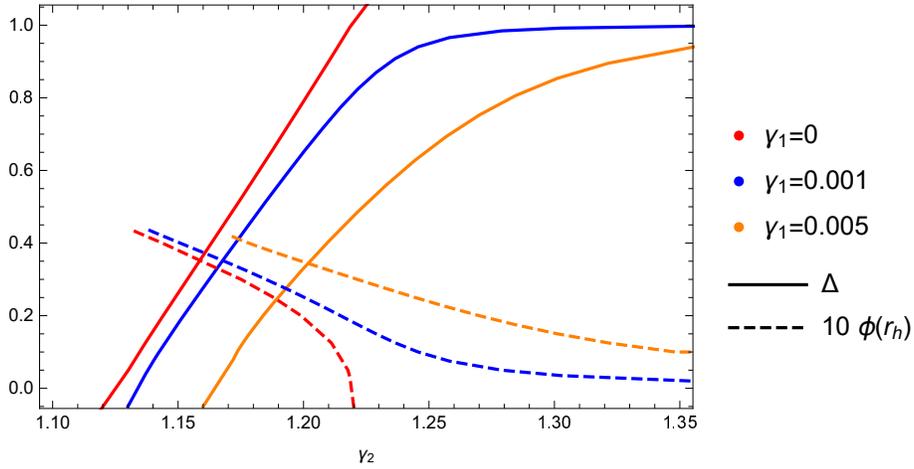}}
\end{center}
\caption{The value $\phi(r_h)$ and the quantity $\Delta$ as function of $\gamma_2$ 
for several values of $\gamma_1$.
\label{fig_mix_one_node}
}
\end{figure}

In the shift-symmetric case, \textit{i.e.} with  $\gamma_2=0$, the condition (\ref{regular_condition}) 
drastically reduces the spectrum of hairy black holes. For each value of 
$\gamma_1 < 1/\sqrt{96}$ a single solution is allowed whith ``$+$'' sign (since $\phi'(r_h)$ does not depend on $\phi(r_h)$ but only on the fixed parameters $\gamma_1$ and $r_h$) and the scalar field is a monotonic function.
Consequently, excited solutions (\textit{i.e.} with $\phi(r)$ presenting  nodes) do not occur. 
By contrast, for the spontaneously scalarized black holes (\textit{i.e.} with $\gamma_1 = 0$), the linear equation (\ref{klein_gordon})  possesses -- in principle -- a  
series of critical values of $\gamma_2$ corresponding to 
normalizable eigenfunctions $\phi(r)$ presenting one or more nodes. 
Any of these solutions leads to a branch of excited hairy black holes  
of the coupled system ($\gamma_1 \neq 0, \gamma_2 \neq 0$). 
We constructed numerically the branch corresponding to the first excited (or one-node) solution.
 Values $\Delta$ and $\phi(x_h)$ are reported  on Fig. \ref{fig_mix_one_node}
as functions of $\gamma_2$ for a few values of $\gamma_1$
 (the red lines correspond to $\gamma_1=0$).
 As for the fundamental (or no-node) solution discussed above, we see that the first excited hairy black holes 
exists for $\gamma_2 \in [\gamma_{2,c}, \gamma_{2,M}]$ where the lower (resp. upper) bound of this interval
corresponds to  $\Delta=0$  (resp. to the second eigenvalue of \eqref{klein_gordon}).

Switching on the parameter $\gamma_1$ leads to a deformation of these excited hairy black holes.
The results of  Fig. \ref{fig_mix_one_node} suggest that the excited black holes exist only for $\gamma_2 \geq \gamma_{2,c}$.
This  contrasts drastically with the spectrum of fundamental solutions (see Fig.\ref{fig_mix_1}). 
It is tempting to say that the fundamental  solutions  are ``attracted'' by the SZ solutions occurring in the $\gamma_2=0$ limit.
Having no equivalent, the excited solutions exist only for large values of $\gamma_2$.
\subsubsection{Influence of a mass term}
\label{massterm}
In the previous section, the scalar field $\phi$ was supposed to be massless.
In this section, we discuss the effect  of a massive scalar field 
on the spectrum of hairy black holes. For simplicity we restrict the presentation
to the spontaneously scalarized solutions  -- \textit{i.e.} setting $\gamma_1=0$ -- and to 
the mass term only in the potential (\ref{potential}) -- \textit{i.e.} $\lambda_4=\lambda_6 = 0$. 

\begin{figure}[h!]
\begin{center}
{\label{non_rot_cc_1}\includegraphics[width=12cm]{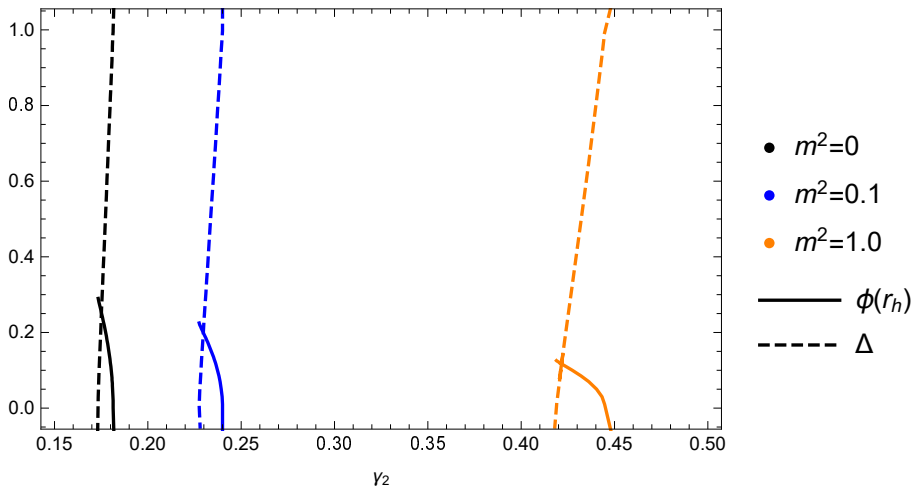}}
\end{center}
\caption{The value $\phi'(r_h)$ and the quantity $\Delta$ as function of $\gamma_2$ 
for several values of $\gamma_1$ for 
the solutions with $r_h=1$.
\label{fig_mix_mass}
}
\end{figure}

In the case of a massive scalar field, the regularity condition (\ref{regular_condition}) is more involved~:
\be
      \phi'(r_h) = \frac{-B \pm \sqrt{\Delta}}{2 A}
\ee
with
\be
     A = -\phi_0(12 \gamma_2 - m^2 r_h^2(r_h^2 + 8 \gamma_2 \phi_0^2) + 4 \gamma_2  r_h^4 \phi_0^4) \ \ , \ \ 
		 B = 8 \gamma_2 \phi_0 (r_h^2 - \phi_0^2 (r_h^4 + 8 \gamma_2 r_h^2 - 64 \phi_0^2 \gamma_2^2)) \ \ , \ \ 
\ee
\be
     \Delta = (1 - m^2 \phi_0^2 r_h^2 )^2 \biggl( 		
		r_h^2(r_h^4 - 384 \gamma_2^2 \phi_0^2)  
		+ 256 m^2 \gamma_2^2 \phi_0^4 (r_h^4 + 12 \gamma_2 r_h^2 - 96 \gamma_2^2 \phi_0^2)
		+ 4096 m^4 \gamma_2^4 \phi_0^8 r_h^2
                                           \biggr)
\ee
and  we posed $\phi(r_h) = \phi_0$. The temperature of the black hole can be evaluated by using~:
\be
     N'(r_h) = \frac{1 - m^2 r_h^2 \phi(r_h)^2}{r_h + 4 \phi'(r_h)(\gamma_1 + 2 \gamma_2 \phi(r_h))} \ \
\ee
instead of (\ref{nprime}).
This suggests that hairy black holes occuring from a massive scalar field can eventually be extremal.
However for all values of $m$ that we adressed
(see Fig. \ref{fig_mix_mass}), the parameter $\phi(r_h)$ remains too small for extremal black holes to form.

The numerical results reveals that the inclusion of a massive scalar field results in shifting the
interval of existence in $\gamma_2$ to larger values, as  demonstrated by Fig. \ref{fig_mix_mass}.
The critical phenomena  limiting  the interval of existence is of the same as discussed above. 
The shift to larger values  of the interval of $\gamma_2$  while increasing $m$
can be understood by examining  the field equation of the scalar field. With the 
assumptions made in this section  (for instance~: a real scalar field,  a mass term only and $\gamma_1=0$),
 \eqref{eqphi} reads
\[- \square \phi =   2 \left(\gamma_2\mathcal{I}(g)-m^2\right) \phi.\]
One can see that the mass act as a ``negative shift constant'' on the term $\gamma_2 \mathcal{I}(g)$. 
For $m = 0$ the scalarised solutions  appears only when the Gauss-Bonnet term becomes sufficiently important (\emph{\textit{i.e.}} when $\gamma_2$ is large enough to ensure the term $\mathcal{I}(g)$ to  trigger the scalar field). It is then  intuitive to assert that, since the mass just shift down the trigger term of the scalar field, higher values of $\gamma_2$ are needed to allow for spontaneous
scalarisation.

\section{Boson stars}
As we mentionned in Sect. 2, it is well known (see e.g. \cite{jetzer})
 that regular solutions -- boson stars -- exist within a large subclass of the lagrangian (\ref{action}).
Let us first specify the conditions~:
\begin{itemize}
\item The scalar field is complex, of the form \eqref{scalar} with $\omega \neq 0$.
Accordingly, the Lagrangian possesses a $U(1)$-global symmetry.
\item The linear coupling to the Gauss-Bonnet term will be set to zero so $\gamma_1=0$ in (\ref{f_function}).
This is because we want to limit ourselves to a polynomial lagrangian in $\phi$.
\item The potential should contain at least a mass term, so $m > 0$ in (\ref{potential}).
\end{itemize}
 Asymptotically, the functions $N(r), \phi(r)$ behave according to
\be
\label{asymp_bs}
     N(r) = 1 - \frac{2M}{8 \pi r}  + O(1/r^2) \ \ \ , \ \ \ \phi(r) \propto \exp(- r \sqrt{m^2 - \omega^2} ) \ \ ,
\ee
contrasting with (\ref{mass_charge}).
The exponential decay of the scalar field  demonstrates the  crucial role of frequency parameter $\omega$ and  of the mass $m$; 
in particular boson stars exist for $\omega < m$.
Beside the mass $M$, the solutions are further characterized by the Noether charge 
associated to  the U(1)-symmetry of the Lagrangian.
The Noether current and the  calculation of the charge $Q$ can be found in numerous papers
(see e.g.\cite{Kleihaus:2005me}), for brevity we give the final form
of the integral to be computed to evaluate the charge~:
\be
\label{noether_charge}
        Q =  8 \pi \omega \int  \frac{r^2 \phi^2}{N \sigma} \diff r.
\ee
This quantity is interpreted 
as the number of elementary bosons of mass $m$ constituting the star
\footnote{One formal analogy can be made with the total charge $Q_{EM}$ 
of a system of $N$ particles of electric charge $q$. In such a case 
 the number of components is obtained via the relation $N = Q_{EM}/q$.}.

The construction of boson stars is achieved by solving the field equations of Sect. 2.2
for $r \in [0,\infty]$. The regularity at the origin, the asymptotic flatness and 
the localization of the scalar field imply the following set of boundary conditions~:
\be
      N(0) = 1 \ \ , \ \ \phi(0) = F_0 \ \ , \ \ \phi'(0) = 0  \ \ , \ \ A(r \to \infty) = 1 \ \ , \ \ \phi(r \to \infty) = 0 \ 
\ee 
which determine the boundary value problem.
Practically, the value $F_0$ at the center is used as control parameter in the numerical resolution;
the frequency $\omega$ has to be fine-tuned as a function of $F_0$ for all boundary conditions to be obeyed. 
The frequency $\omega$, the mass $M$ and the  Noether charge $Q$  can then be evaluated as functions of $F_0$.
\subsection{Solutions without self-interaction}
\begin{figure}[h!]
\begin{center}
{\label{non_rot_cc_1}\includegraphics[width=12cm]{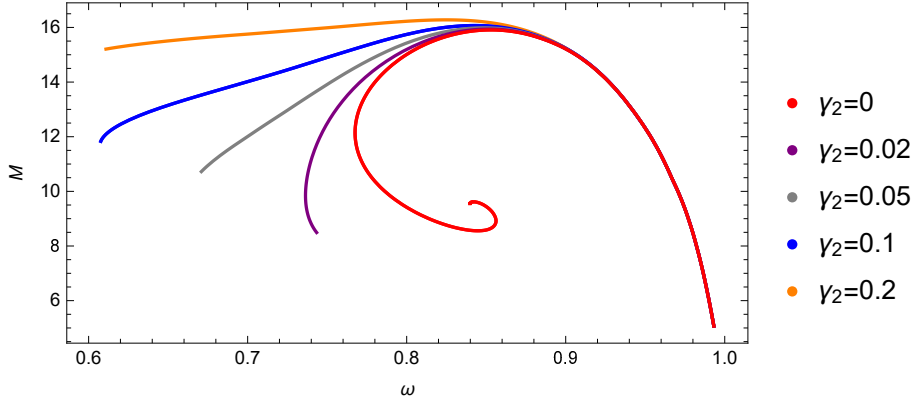}}
\end{center}
\caption{The mass of the boson star as a function $\omega$ for no-selfinteraction and for different values of $\gamma_2$. 
\label{fig_boson_star_mc}
}
\end{figure}

Let us  first discuss the solutions for a pure mass potential (\textit{i.e.} for $\lambda_4 = \lambda_6 = 0$ in (\ref{potential})).
The minimally coupled boson stars (\textit{i.e.} corresponding to $\gamma_2 = 0$)  
exist on a finite interval of the frequency $\omega/m$, that is to say
for $\omega/m \in [\omega_{\min}/m,1.0]$ with $\omega_{\min}/m \approx 0.76$. 
The plot of the mass $M$ versus $\omega$ 
presents the form of a spiral as shown by the red line on Fig. \ref{fig_boson_star_mc}.
From this pattern, it results that two or more solutions can exist with the same frequency 
on specific sub-intervals  of $\omega$. The vacuum (\textit{i.e.} Minkowski space-time) is approached 
for $F_0 \to 0$ which coincides with the limit $\omega/m \to 1$.
The  phenomenon limiting the boson stars in the center of the spiral is the following~:
while increasing $F_0$   the effects of gravity get stronger at the center of the lump, in particular  
 the value $\sigma(0)$ decreases, finally approaching  zero. 
Correspondingly the value $R(0)$ of the Ricci scalar gets arbitrarily large and 
a configuration with a singularity at the center is approached.

\begin{figure}[h!]
\begin{center}
{\label{non_rot_cc_1}\includegraphics[width=8cm]{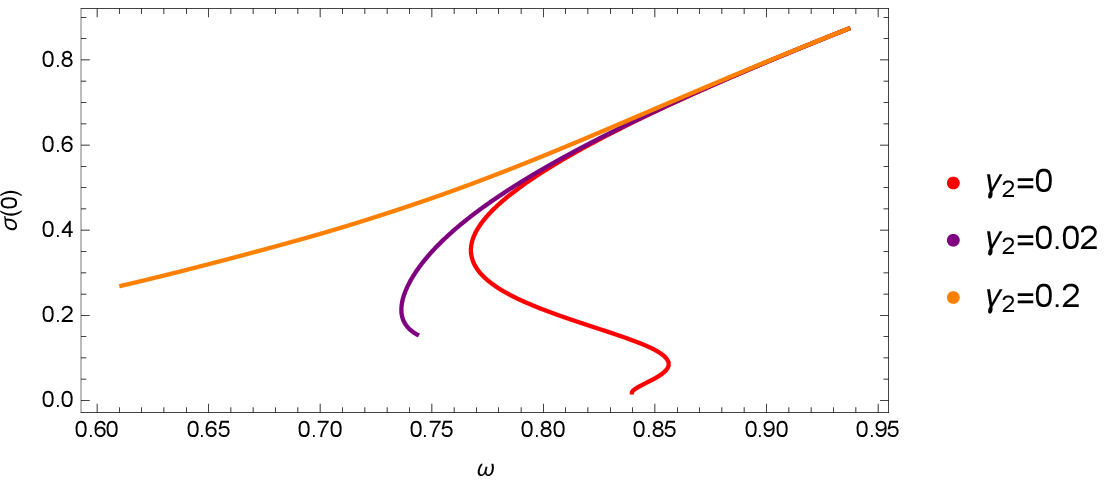}}
{\label{non_rot_cc_2}\includegraphics[width=8cm]{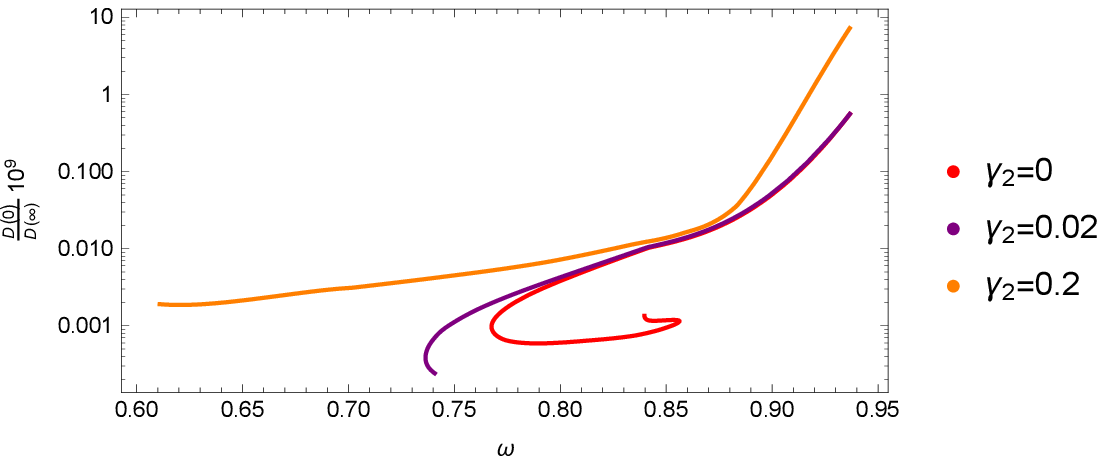}}
\end{center}
\caption{The value $\sigma(0)$ as function of $\omega$ for boson stars and three values of $\gamma_2$.
Right~: Idem for the discriminant of the system of equations.
\label{critical_limits}
}
\end{figure}

We now discuss  the influence of the non-minimal coupling (\textit{i.e.} with $\gamma_2 > 0$) on this pattern.
A look at Fig. \ref{fig_boson_star_mc} reveals that the 
$M$ versus $\omega$ curve has the tendency to unwind for $\gamma_2 > 0$ and that the boson stars 
exist on a larger interval of  $\omega$.
The nature of the phenomenon limiting the curves corresponding to $\gamma_2 > 0$ on Fig. \ref{fig_boson_star_mc}   
is different from  the case $\gamma_2 = 0$ mentionned above.
Denoting $D(r)$ the denominator of the function $F_3$ in (\ref{radial_equations}),
it turns out that the values $\sigma(0)$, $D(0)$ both decrease when  $F_0$ increases.
However the numerical results strongly indicate that  $D(0)$ tends to zero much quicker than $\sigma(0)$ once $\gamma_2 > 0$. 
This statement is hard to demonstrate because the numerical integration of the equations becomes particularly tricky in this limit.
Within the coordinate system used
both the numerator and denominator entering  in $F_3$ become quite large in a region of the interval
of integration and the accuracy of the numerical solution get lost.
The situation is illustrated on Fig. \ref{critical_limits} where the pattern of the solutions 
is shown in the $\omega-\sigma(0)$ plane (left-figure) and in the $\omega-\frac{D(0)}{D(\infty)} 10^9$ plane (right-figure).
In this  plot, the quantity $D(r)$ has been normalized  with respect 
to   $D(\infty)$ in order to compare the curves for the different values of $\gamma_2$ considered.
The logarithmic scale used on the vertical axis of the right plot illustrates the huge variation 
of  $D(r)$ while approaching the critical configuration.
The two plots confirm  that for, $\gamma_2 \neq 0$, 
the limit of existence of the boson stars 
is related to the behaviour of $D(0)$, rather than to $\sigma(0)$ 
whose values remain finite.

Note that the unwinding phenomenon of the $\omega-M$ relation seems to be closely related to the Gauss-Bonnet term.
It was first observed in the construction of
 boson stars in Einstein-Gauss-Bonnet  gravity in five dimensions \cite{Hartmann:2013tca}. 
 In this case, the Gauss-Bonnet term is fully dynamic and does not need coupling to extra field.
\subsection{Effect of a self-interacting term}
Because the self-interacting potential depends on two independant parameters
(namely $\lambda_4, \lambda_6$), we limited  the investigation to the  
potential  of the form~: $V = \phi^2(1 - \phi^2)^2$. Presenting three degenerate vacua ($\phi=0, \pm1$),
this potential offers rich  possibilities for topological solitons \cite{lohe}. 
Recently  it was used in \cite{Dorey:2011yw} for the study of 
kink-anti-kink collisions in 1+1 dimensions and in \cite{Brihaye:2015veu}
to study  boson stars in 3+1 dimensions.

\begin{figure}[h!]
\begin{center}
{\label{non_rot_cc_2}\includegraphics[width=12cm]{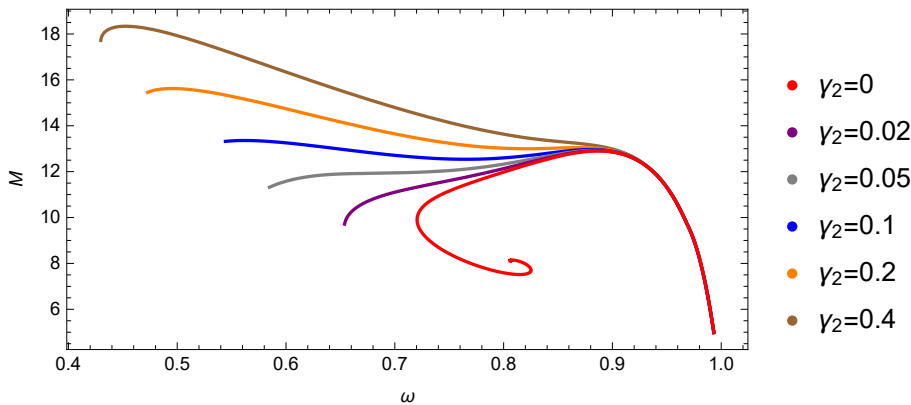}}
\end{center}
\caption{The mass of the boson star as a function $\omega$ for a self-interacting potential.
\label{fig_boson_star_si}
}
\end{figure} 

The general effect  of the self-interacting potential on the solutions
is that the interval of  frequencies of the boson stars is significantly larger that for the mass potential.
Especially the minimal value $\omega_{\min}$ is systematically lower (e.g. $\omega_{\min} \approx 0.72$ for $\gamma_2=0$).
These features 
are illustrated by Fig. \ref{fig_boson_star_si}, to be compared with Fig. \ref{fig_boson_star_mc}.
The unwinding   feature of the mass-frequency graphic occurring for the mass potential 
also takes place when the self-interaction is present. The minimal value $\omega_{\min}$ is again systematically lower although remaining strictly positive. 
The combination of self-interacting scalar field and non-minimal coupling to the Gauss-Bonnet term is therefore
not suitable to allow for purely real soliton solutions.

One can also note that, in the presence of the self-interaction, an increase of $\gamma_2$ can lead to solutions with drastically higher mass compared to  solutions with $\gamma_2 = 0$ (compare, for exemple, the difference between the curves corresponding to $\gamma_2 = 0$ (red) and $\gamma_2 = 0.2$ (orange) on Fig. \ref{fig_boson_star_mc} and Fig. \ref{fig_boson_star_si}).

\subsection{Classical stability}
We now  address the stability of the boson stars  by invoking a ``classical argument''.
With the interpretation of $Q$ as the number of bosons of mass $m$ in the lump, it is natural to compare the quantity $m Q$  to the total mass of the solution $M$. 
If $M < m Q$, the total mass of the boson star is lower than the sum of its components, 
\textit{i.e.} the total energy of the system is lower than the energy corresponding 
to $Q$ ``free'' bosons. In such a case, as for the mass defect in atoms, we will say 
that the system is stable, in the sense that the $Q$ bosons can't exist in a ``free'' 
form but have to be bounded within the star.
Following the same lines, the case $M > m Q$ will correspond to unstable configurations
(remember $m=1$ as fixing our scale, see section \ref{rescale}). 

\begin{figure}[h!]
\begin{center}
{\includegraphics[width=8cm]{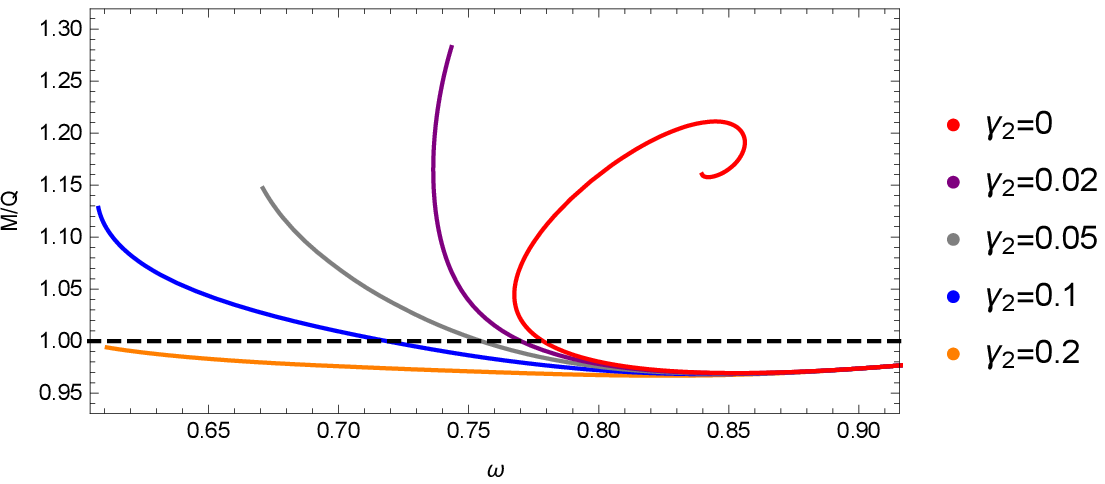}}
{\includegraphics[width=8cm]{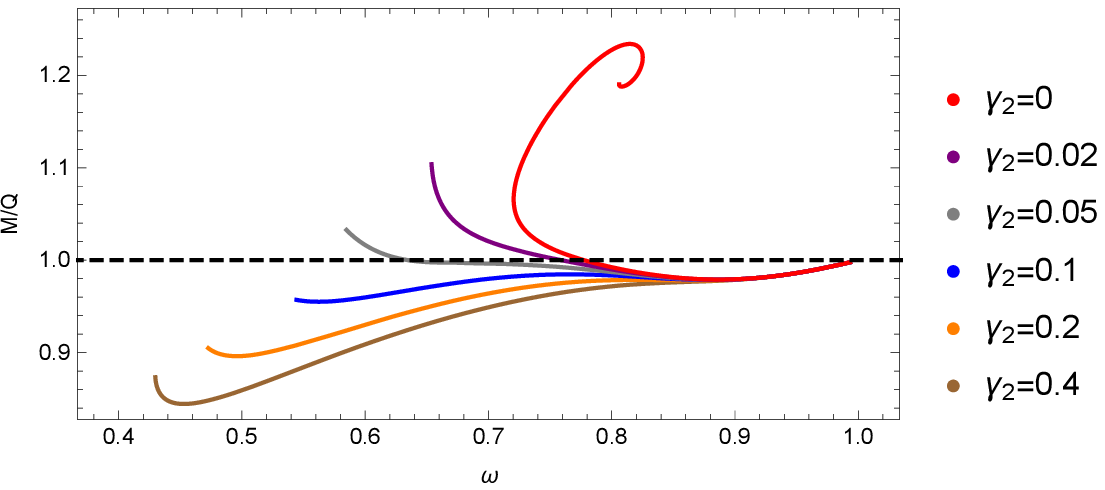}}
\end{center}
\caption{Left : The quantity $M/Q$ 
  as function of $\omega$ for several values of $\gamma_2$ for solutions without self-interaction. Right : Idem for  self-interacting solutions.}
\label{figstab}
\end{figure}

The quantity $M/Q$ is reported as a function of $\omega$  
 on Fig. \ref{figstab}  for several values of $\gamma_2$. 
The left part of Fig. \ref{figstab} characterizes solutions with the mass term only. 
We see that the solutions emerging from the vacuum limit (\textit{i.e.} $\omega/m=1$) are classically stable
and remain so for sufficiently 
high values of $\omega$, say for $\omega \ge \omega_{s}$ where  $\omega_{s}$ is 
such that $M/Q = 1$. For values of $\omega$  such that several 
solutions coexist, the most massive is the most stable.

The most interesting result concern the influence of $\gamma_2$ on this pattern.
 As one can see on the plot, $\omega_s$ decreases when $\gamma_2$ increases while, 
for fixed $\omega$, $M/Q$ decreases when $\gamma_2$ increases. 
Consequently, the presence of the interaction between the scalar 
field and the geometry enhance the stability of the solutions.

This feature remains qualitatively the same for self-interacting solutions 
as illustrated on the right part of the figure. 
The presence of the self-interaction reinforces the effects of the curvature 
and the boson stars are even more stable compared to non-self-interacting ones.
For sufficently high values of $\gamma_2$ (say for $\gamma_2 > 0.075$), 
the whole set of solutions is classically stable.

\section{Conclusion}
The investigation for hairy black holes in  gravity extended by a Gauss-Bonnet term coupled to a scalar field
was a source of huge activity over the past years. 
In particular the stability of such objects was examined in details in
 \cite{Minamitsuji:2018xde},\cite{Silva:2018qhn};
 the construction of such black holes in the presence of a cosmological constant was reported in 
 \cite{Bakopoulos:2018nui}.
The coupling function of the scalar field to the Gauss-Bonnet term is, up to now, left as an 
arbitrary freedom but its form  lead to different patterns for the solutions
  and turns out to be important for the  stability of the hairy black holes.

In this paper we considered as coupling a superposition of the linear and quadratic powers of the scalar field. 
While spontaneoulsy scalarized black holes -- with purely quadratic coupling constant $\gamma_2$ -- appear on a very limited interval of the coupling constant $\gamma_2$, we showed that, when adding a linear part 
(even with small coupling $\gamma_1$), two branches of hairy black holes exist. 
One of these branches  is very close to the spontaneously scalarized black holes while the second extend backward
to a solution with shift-symmetric scalar field.
This feature is specific for the fundamental solutions and is not repeated for excited solution (\textit{i.e.} with scalar field
presenting  nodes) .

Extending the scalar sector of scalar-tensor gravity to a massive, complex field, we were also able to construct boson star solutions
in the full theory. The qualitative and quantitative effects of the  Gauss-Bonnet term have been reported in details in Sect. 4 revealing, for instance, that the presence of the quadratic coupling constant $\gamma_2$ can drastically increase the maximal mass of these objects and the range of $\omega$ (the frequency of the complex scalar field) for which these solutions exist. In this context, we also show that the critical phenomenon limiting the existence of solutions is different in the minimally and non-minimally coupled case. Interestingly, our results demonstrate that the coupling to the Gauss-Bonnet invariant and/or the inclusion of a self-interacting
potential of the scalar field enhances the domain of classical stability of the boson stars.

Finally, in the Appendix we studied the solutions for scalar-tensor gravity extended by the same kind of coupling
of the scalar field to the Chern-Simons invariant.
Here the space-time is endowed with a NUT charge. The pattern of Nutty-Hairy black holes
is qualitatively  similar to the Gauss-Bonnet case.


\appendix
\makeatletter
\let \@standardsection \thesection
\def\thesection{Appendix \@standardsection}
\makeatother

\section{\hspace{-3mm}: Coupling to the Chern-Simons invariant}
\label{appendix}

In this appendix, we provide an analysis of hairy black holes in the model 
\eqref{lagrangian} where the curvature invariant is chosen as the Chern-Simons-scalar : 
\begin{equation*}
{\cal I}(g)=\mathcal{L}_{CS}\equiv R \tilde R = {\,^\ast\!}R^a{}_b{}^{cd} R^b{}_{acd} \ \ , \ \ 
{^\ast}R^a{}_b{}^{cd}\equiv \frac12 \eta^{cdef}R^a{}_{bef}\,
\end{equation*}
where
  ${\,^\ast\!}R^a{}_b{}^{cd}$ is the Hodge dual of the Riemann-tensor,
 $\eta^{cdef}=\epsilon^{cdef}/\sqrt{-g}$  the 4-dimensional Levi-Civita tensor 
			and $\epsilon^{cdef}$ the Levi-Civita tensor density.

The construction of classical solutions  with a non-trivial Chern-Simons term
can be performed by enforcing rotations in the metric or by endowing Space-Time with  a NUT charge.
Nutty-Hairy black holes in Einstein-Chern-Simons gravity were 
 constructed   in  \cite{Brihaye:2005ak} and \cite{Brihaye:2018bgc} for $\gamma_2=0$ and $\gamma_1=0$ respectively. 
Similar solutions within Einstein-Gauss-Bonnet (rather than Chern-Simons) gravity were obtained in 
\cite{Brandelet:2017nbc}.
The field equations are  given by \eqref{eqG} and \eqref{eqphi} 
with a  different  expression of $T_{\mu\nu}^{(\mathcal{I})}$ which  can be found in \cite{Brihaye:2018bgc} 
with the same  notations as in  Sect. \ref{action}.

\subsection*{The ansatz}

To construct the solutions we use a metric of the form 
%
\begin{eqnarray*}
\diff s^2=-N(r)\sigma^2(r)(\diff t + 2n\cos \theta  \diff \varphi)^2+\frac{\diff r^2}{N(r)}+g(r)(\diff \theta^2+\sin^2 \theta \diff\varphi^2)~,
\end{eqnarray*}  
generalizing the Schwarzschild-NUT solution. 
Here $\theta$ and $\varphi$ are the standard angles on $S^2$ with the usual range while $r$ and $t$ are the ``radial'' and ``time'' coordinates respectivelly.
The NUT parameter $n$ appears as a  coefficient in the differential form
$\diff t+2n\cos \theta  \diff\varphi$ (note that $n\geqslant 0$, without any loss of generality).
 When evaluated with this metric, the Chern-Simons density
$\mathcal{L}_{CS}$ is actually proportional to the NUT charge; 
so it vanishes identically for spherically symmetric solutions ($n = 0$) but becomes 
non trivial for $n \neq 0$, ensuring a non-trivial behaviour of the scalar 
field via the curvature induced scalarization 
only for $n \neq 0$.

In the decoupling limit $\gamma_1=\gamma_2 = 0$ (implying $\phi=0$), the functions $N(r), \sigma(r)$ and  $g(r)$
are known explicitely~: 
\begin{eqnarray*}
\label{TN}
N(r)=1-\frac{2(M r+n^2)}{r^2+n^2}\ , \qquad \sigma(r)=1 \ , \qquad g(r)=r^2+n^2  \ .
\end{eqnarray*}
This metric therefore possesses an  horizon  at
\begin{eqnarray*}
\label{rH}
 r_h=M+\sqrt{M^2+n^2}>0.
\end{eqnarray*} 
As in the Schwarzschild limit, $N(r_h) = 0$ is only a coordinate singularity where all curvature invariants are finite. 
In fact, a nonsingular extension across this null surface can be found~\cite{Hawking:1973uf}.
Completing the metric (\ref{TN}), the ansatz for the scalar field is the same as Eq. \eqref{scalar}.

\subsection*{Numerical results}

In the same spirit as in the main part,
we have constructed the black hole solutions in the Einstein-Chern-Simons (ECS) model  with the mixed
coupling (\ref{lagrangian}) and using a Nutty space-time in order to make the Chern-Simons term non trivial.

For generic values of $\gamma_1, \gamma_2$, no explicit solution can be found and, again, we relied on a numerical
technique.  For the construction, we used the gauge $\sigma(r)=1$.
Then the Einstein-Chern-Simons equations can be transformed into a system of three coupled differential
equations of the second order for the functions $N(r), g(r)$ and $\phi(r)$.
The desired asymptotic form of the solutions require
\begin{equation*}
        N(r\to \infty) = 1  \ \ , \ \ \sigma(r \to \infty) = 1 \ \ , \ \ \phi(r\to \infty) = \frac{Q_s}{r}\
\end{equation*} 
where $Q_s$ is the scalar charge.
Imposing an horizon $r=r_h$, \textit{i.e.} $N(r_h) = 0$, the conditions of regularity of the solution at the horizon 
can be determined on the first few coefficients of the Taylor expansion~:
\begin{equation*}
     N(r) = N_1(r-r_h) + O((r-r_h)^2) \ \ , \ \ 
		g(r) = g_0 + g_1(r-r_h)+ O((r-r_h)^2) \ \ , \ \ 
		\phi(r_h) = \phi_0 + \phi_1 (r-r_h) + O((r-r_h)^2)
\end{equation*}
Two conditions are finally necessary~:   
\begin{equation*}
      g'(r_h) = \frac{1}{N_1}(2 - g_0\phi_0 m^2 - 2 n N_1 \phi_1 (\gamma_1 + 2 \gamma_2 \phi_0)) \ \ ,
\end{equation*}
\begin{equation*}
    24 \gamma_2 \phi_0^2 \phi_1 (N_1)^3 
		+ N_1 (2 \gamma_2 n g_0m^2 \phi_0^3 - 12 \gamma_2 n \phi_0 - g_0^2 \phi_1)
		+ g_0^2 \phi_0 m^2 = 0.
\end{equation*}

\begin{figure}[h!]
\begin{center}
{\label{non_rot_cc_1}\includegraphics[width=12cm]{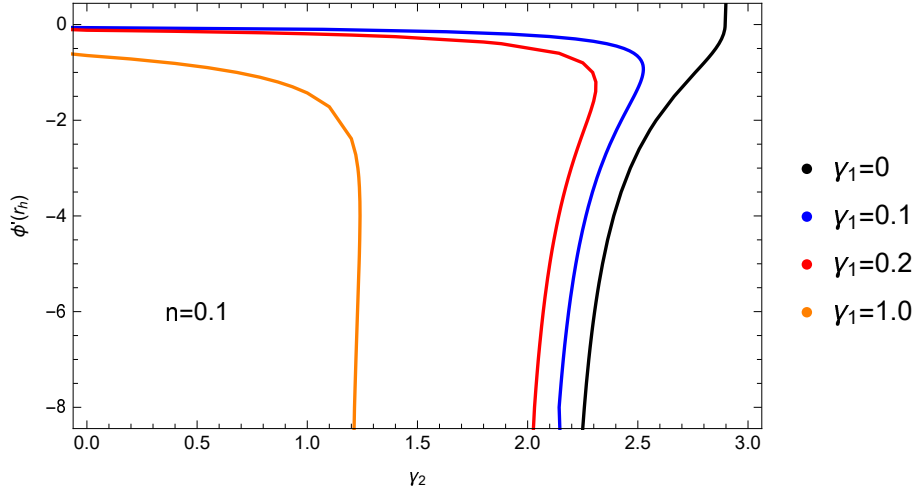}}
\end{center}
\caption{The value $\phi'(r_h)$  as function of $\gamma_2$ for several values of $\gamma_1$ for 
the solutions with $r_h=1$ and $n=0.1$.
\label{fig_ecs}
}
\end{figure}

The pattern of the solutions found for the ECS case is very similar to the case of EGB. 
In particular, the solutions available for non-zero values of $\gamma_1, \gamma_2$ 
smoothly extrapolate between the limits $\gamma_1=0$ and $\gamma_2=0$ found in 
\cite{Brihaye:2018bgc} and \cite{Brihaye:2005ak}.
The results are summarized on Fig. \ref{fig_ecs} for $n=0.1$ but we have checked that 
they  are qualitatively similar for different values of $n$.
\clearpage


 \end{document}